\documentclass[prd,aps,nofootinbib,floatfix,11pt]{revtex4}
\usepackage{amsmath,graphicx,epsfig,amssymb,dsfont,mathtools,}
\usepackage[usenames]{color}
\usepackage{ulem} 
\usepackage{bigstrut}
\usepackage{slashed}
\usepackage{multirow}
\usepackage{subfigure}

\allowdisplaybreaks

\begin{document}\title{Chiral dynamics and S-wave contributions in $\bar B^0/D^0 \to \pi^{\pm}\eta\  l^{\mp} \nu$
  decays }

\author{Yu-Ji Shi$^{1}$~\footnote{Email:shiyuji92@126.com}
  and Ulf-G. Mei{\ss}ner~$^{1,2,3}$~\footnote{Email:meissner@hiskp.uni-bonn.de}}
\affiliation{$^1$ Helmholtz-Institut f\"ur Strahlen- und Kernphysik and Bethe Center \\ for Theoretical Physics,
  Universit\"at Bonn, 53115 Bonn, Germany\\
  $^2$ Institute for Advanced Simulation, Institut f{\"u}r Kernphysik and J\"ulich Center for Hadron Physics,
  Forschungszentrum J{\"u}lich, D-52425 J{\"u}lich, Germany\\
$^3$ Tbilisi State University, 0186 Tbilisi, Georgia}

\begin{abstract}
In this work, we analyze the semi-leptonic decays $\bar B^0/D^0 \to (a_0(980)^{\pm}\to\pi^{\pm}\eta)  l^{\mp} \nu$
within light-cone sum rules. The two and three-body light-cone distribution amplitudes (LCDAs) of the $B$
meson and the only available two-body LCDA of the $D$ meson are used. To include the finite-width effect of
the $a_0(980)$, we use a scalar form factor to describe the final-state interaction between the $\pi\eta$ mesons,
which was previously calculated within unitarized Chiral Perturbation Theory.  The result for the decay
branching fraction of the $D^0$ decay is in good agreement with that measured by the BESIII Collaboration,
while the branching fraction of the $\bar B^0$ decay can be tested in future experiments.
\end{abstract}
\maketitle

\section{Introduction}
The study of scalar mesons is an interesting topic in the hadron physics, partly due to the fact that
the scalar mesons have the same quantum numbers as the QCD vacuum. Furthermore, the highly non-perturbative
nature of the strong interactions at low energies makes it difficult to understand the internal structure and
dynamics of these particles. There are a number of scenarios to understand the structure of the isovector
scalar meson $a_0(980)$, like a $\bar q q$ state, a four-quark state, a molecular state, the glueball picture,
or a hybrid states  \cite{Cheng:2005nb, Weinstein:1982gc, Weinstein:1983gd, Weinstein:1990gu, Achasov:1996ei,Abdel-Rehim:2014zwa,Branz:2007xp,Amsler:1995td, Amsler:1995tu, Amsler:2002ey,Baru:2003qq,Oller:1997ti,Gorishnii:1983zi,Sun:2010nv,Pelaez:2003dy,Hooft:2008we,Dai:2018fmx,Maiani:2004uc}. However, until now there is no definite  conclusion on
which scenario is correct, however, there are some indications in favor of the molecular picture.

The semi-leptonic heavy meson decays are ideal platforms for the study of scalar mesons such as the $a_0(980)$.
Experimentally, such processes have a much cleaner background than e.g. hadronic decays, and theoretically, all
the strong dynamics is encapsulated in the hadronic transition matrix elements, which are the objects to be dealt
with. Nowadays there are a number of theoretical investigations for the $B\to a_0(980)$ \cite{Issadykov:2015iba,Cheng:2013fba,Cheng:2003sm,Wang:2008da,Liang:2019eur,Chen:2021dwn} and the $D\to a_0(980)$ transitions \cite{Cheng:2017fkw,Soni:2020sgn,Huang:2021owr,Ahasov:2019ome,Maiani:2007iw}. In the narrow-width limit, where the $a_0(980)$ is
considered as a quasi-stable particle, calculating these single-body transitions is enough for 
predicting the decay branching fractions. However, in real physical processes the $a_0(980)$ is an intermediate
resonance which can further decay into light mesons. Recently, the BESIII Collaboration has announced a
measurement of the branching fraction \cite{Ablikim:2018ffp}:
\begin{align}
{\cal B}(D^0 \to (a_0(980)\to\pi^{-}\eta)\  e^{+} \nu)=(1.33\pm0.09)\times 10^{-4}.
\end{align}
Note that the actual measured final-state is $\pi\eta$. Thus to analyze the full decay process one must
consider the finite-width effect and should also deal with the final-state interaction between the $\pi$
and the $\eta$.

As shown in the previous works, see e.g. \cite{Doring:2013wka,Shi:2015kha,Shi:2017pgh}, the S-wave two-meson
final-state interaction can be described by a two-meson scalar form factor. This is justified by the
Watson--Migdal theorem~\cite{Watson:1952ji,Migdal:1956tc}, which ensures that the phase of the
$B/D\to\pi\eta$ transition matrix element in the semi-leptonic decay must be equal to the phase of
the $\pi\eta$ elastic scattering amplitude. The two-meson scalar form factor was calculated using
unitarized Chiral Perturbation Theory (uChPT) assisted by a numerical iteration based on a
dispersion relation \cite{Shi:2020rkz}. Such a scalar form factor satisfies the unitary constraint and is
free from unphysical singularities (which often appear in the framework of uCHPT, see e.g. the discussion
in Ref.~\cite{Du:2017ttu}). In those previous works the transition matrix element was calculated from
Light-Cone Sum Rules (LCSRs), where the light-cone distribution amplitudes (LCDAs) of the scalar meson
or the generalized  LCDAs for the final two mesons are used. In this work, instead of assuming the structure
of the $a_0(980)$ and test it with the corresponding decay width calculation, we aim to theoretically
reproduce the measured branching fraction of $D^0 \to (a_0(980)\to\pi^{-}\eta)  e^{+} \nu$. Therefore,
we choose to use the LCDAs of the initial heavy meson and create the $\pi\eta$ state by interpolating a
scalar current with the same quantum numbers on the vacuum. For the detailed calculation procedure
we follow Refs.~\cite{Cheng:2017smj,Cheng:2019tgh}, where the transitions form factors for
$B\to(\rho\to\pi\pi)$ and $B_s\to(f_0(980)\to KK)$ were calculated.

This article is organized as follows: Section~II is an illustration on the framework of the LCSR, where
we derive the sum rule equation for the $\bar B^0/D^0\to \pi\eta$ form factors by considering the
finite-width effect. Section~III gives the numerical results including the form factors and the decay
branching fractions. Section~IV contains a brief summary. Some technicalities are relegated to the
appendices.

\section{$\bar B^0/D^0 \to \pi^{\pm}\eta$ Form Factors in a LCSR}
\label{sec:lc_sum_rules}
In the spirit of LCSRs, to study the transition $\bar B^0/D^0 \to \pi^{\pm}\eta$, one should start
with a correlation function. In the case of a $B^0$ decay with the $b\to u$ transition it reads
\begin{align}
\Pi_{\mu}(p,q) & =i\int d^{4}x\ e^{ip\cdot x}\langle0|T\left\{ J^{\bar{d}u}(x)J_{\mu}^{V-A}(0)\right\} |B^{0}(p+q)\rangle~,
\label{CorrFunc}
\end{align}
where $J^{\bar{d}u}$ is the scalar current with the same quantum number as the final S-wave $\pi^+\eta$ state,
and $J_{\mu}^{V-A}$ is the standard $V-A$ current. Their explicit forms are
\begin{equation}
J^{\bar{d}u}(x)=\bar{d}(x)u(x),\ \ \ \ J_{\mu}^{V-A}(0)=\bar{u}(0)\gamma_{\mu}(1-\gamma_{5})b(0).
\end{equation}
For the case of a $D^0$ decay with the $c\to d$ transition, the correlation function is similar except
that $J^{\bar{d}u}$ and $ J_{\mu}^{V-A}$ should be changed to $J^{\bar{u}d}=\bar u d$ and $ J_{\mu}^{V-A}=\bar d
\gamma_{\mu}(1-\gamma_5)c$, respectively. The correlation function in Eq.~(\ref{CorrFunc}) will be calculated
both at the hadron as well as the quark-gluon level, and the results from these two distinct approaches
are related by  quark-hadron duality. 

\subsection{Hadron Level}
\label{sec:hadronLevel}
\begin{figure}
\begin{center}
\includegraphics[width=0.46\columnwidth]{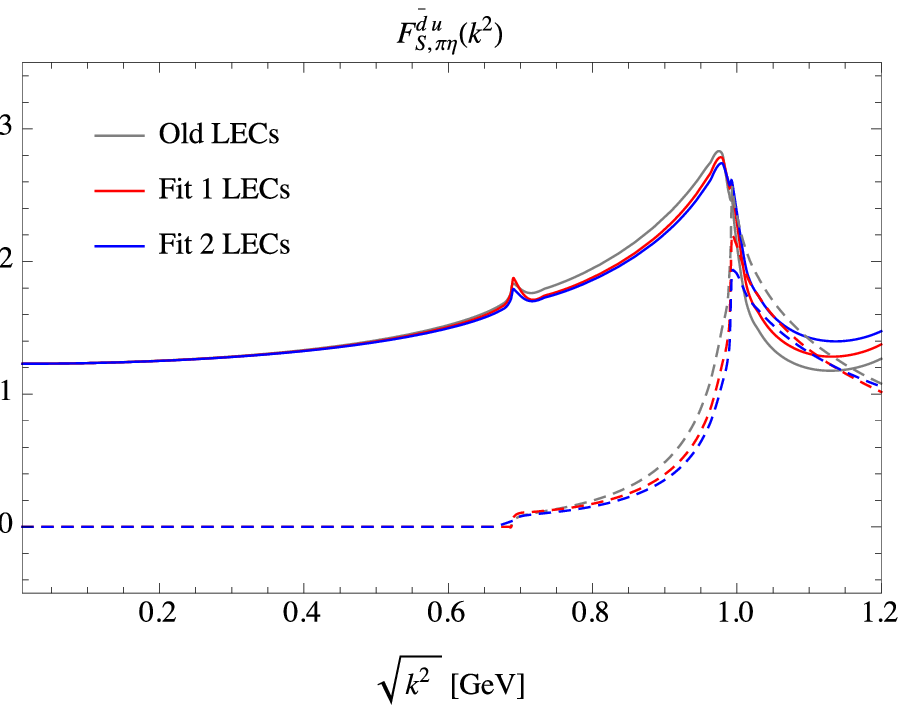} 
\includegraphics[width=0.47\columnwidth]{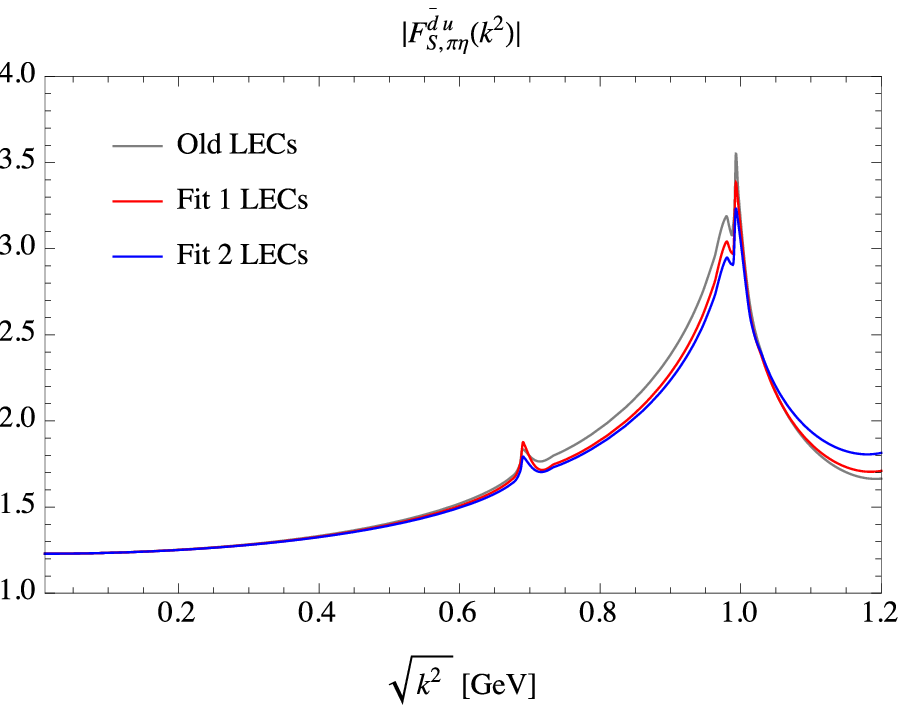} 
\caption{Left panel: Real (red solid lines) and imaginary (blue dashed lines) parts of  $F_{\pi\eta}(k^2)$.
  Right panel: The norm of  $F_{\pi\eta}(k^2)$. For further details, see Ref.~\cite{Shi:2020rkz}.}
\label{fig:pietaFF} 
\end{center}
\end{figure}

In terms of a  dispersion relation, the correlation function in Eq.~(\ref{CorrFunc}) can be expressed as:
\begin{equation}
\Pi_{\mu}^H (p,q)=\frac{1}{\pi}\int_{0}^{\infty}ds\ \frac{{\rm Im}\ \Pi_{\mu}^H (s,q^{2})}{s-p^{2}-i\epsilon}~.
\label{disperRela}
\end{equation}
If the narrow-width approximation is applied, the imaginary part above is just a single meson pole of the
$a_0(980)$ plus a continuous spectrum including higher excited states. However, if the finite-width effect
is considered, the single meson pole should be replaced by a multi-meson state, which has a finite
distribution width. The lowest two-meson state created by $J^{\bar{d}u}$ is $\pi^{\pm}\eta$, while the
heavier two-meson state of $K\bar K$ can be absorbed into the continuous spectrum. The imaginary
part of the correlation function thus reads
\begin{align}
 & 2i\ {\rm Im}\ \Pi_{\mu}^H(s,q^{2})={\rm Disc}\ \Pi_{\mu}^H(s,q^{2})\nonumber \\
  = &\  i\int d\tau_{2}(2\pi)^{4}\delta^{4}(p-k)\langle0|J^{\bar{d}u}(0)|\pi(k_{1})\eta(k_{2})\rangle
  \langle\pi(k_{1})\eta(k_{2})|J_{\mu}^{V-A}(0)|\bar B^{0}(p+q)\rangle+\cdots,\label{CorrFuncIm}
\end{align}
where the ellipsis denotes the higher continuous spectrum contribution, which will be omitted
later for convenience. Further,  $d\tau_2$ is the two-body integral measure:
\begin{equation}
d\tau_{2}=\frac{d^{3}\vec{k}_{1}}{(2\pi)^{3}}\frac{1}{2E_{k_{1}}}\frac{d^{3}\vec{k}_{2}}{(2\pi)^{3}}\frac{1}{2E_{k_{2}}}~,
\end{equation}
with $s=p^2$ and $k=k_1+k_2$. The first matrix element in Eq.~(\ref{CorrFuncIm}) is parameterized by the
two-meson scalar form factor:
\begin{equation}
\langle \pi(k_{1})\eta(k_{2})|J^{\bar{d}u}(0)|0\rangle=B_0 F_{\pi\eta}(k^2)~.\label{Fudpieta}
\end{equation}
Here, $B_0$ is proportional to the QCD quark condensate: $3 F^2 B_0 =-\langle {\bar u} u+\bar d d+\bar s s
\rangle$, with $F$ the pion decay constant in the SU(3) chiral limit. $F_{\pi\eta}(k^2)$ has been calculated in the
Ref.~\cite{Shi:2020rkz} using uChPT associated with a dispersion relation iteration to ensure the unitary
constraint and remove the unphysical singularities. The obtained form factor is applicable up to a
relatively high energy of around 1.2~GeV, and its real and imaginary parts are shown in Fig.~\ref{fig:pietaFF}.
The three lines correspond to the form factors derived by three sets of low-energy constants (LECs) of ChPT.
The one labeled ``old'' is taken from earlier works \cite{Gasser:1983yg,Bijnens:1994ie}.
The other two sets are fitted in Ref.~\cite{Shi:2020rkz} using the latest data, where the authors used
two fit approaches denoted as Fit 1 and Fit 2. In this work, all the three results shown in Fig.~\ref{fig:pietaFF}
will be used, and the difference between them will be taken as the uncertainty of the numerical results.

On the other hand, the second matrix element in Eq.~(\ref{CorrFuncIm}) is parameterized by the
$B^0\to \pi^+\eta$ form factors \cite{Faller:2013dwa}. For the S-wave component, only the axial-vector
current contributes. Its matrix element is parameterized as:
\begin{align}
&-i\left\langle \pi(k_{1}) \eta(k_{2})|\bar{u} \gamma_{\mu} \gamma_{5} b | \bar{B}^{0}(q+k)\right\rangle \nonumber\\
=& F_{t} \frac{q_{\mu}}{\sqrt{q^{2}}}+F_{0} \frac{2 \sqrt{q^{2}}}{\sqrt{\lambda_{B}}}\left(k_{\mu}-\frac{k \cdot q}{q^{2}} q_{\mu}\right) 
+ \frac{F_{\|}}{\sqrt{k^{2}}}\left(\bar{k}_{\mu}-\frac{4(q \cdot k)(q \cdot \bar{k})}{\lambda_{B}} k_{\mu}+\frac{4 k^{2}(q \cdot \bar{k})}{\lambda_{B}} q_{\mu}\right),\label{BtopietaFF}
\end{align}
where $\bar k=k_1-k_2$, $q\cdot k=(1/2)(m_B^2-k^2-q^2)$ and $\lambda_B=m_{B}^{4}+k^{4}+q^{4}-2\left(m_{B}^{2}
k^{2}+m_{B}^{2} q^{2}+k^{2} q^{2}\right)$. Each form factor $F_i$ ($i=0,t,{\|}$) is a function of $q^2$, $k^2$ and
\begin{equation}
q\cdot\bar{k}=(m_{\pi}^{2}-m_{\eta}^{2})\sqrt{\frac{\lambda_{B}}{4k^{2}}+\frac{q^{2}}{k^{2}}}+\frac{\sqrt{\lambda_{B}\lambda_{\pi\eta}(k^2)}}{2k^{2}}{\rm cos}\ \theta_{\pi},
\end{equation}
where $\theta_{\pi}$ is the angle between the pion momenta $\vec k_1$ and $\vec q$ in the rest-frame of
the $\pi-\eta$ system and $\lambda_{\pi\eta}(k^2)=m_{\pi}^{4}+m_{\eta}^{4}+k^{4}-2\left(m_{\pi}^{2}m_{\eta}^{2}
+m_{\eta}^{2} k^{2}+m_{\pi}^{2}k^{2}\right)$.  As illustrated in Ref.\cite{Faller:2013dwa}, one can apply a
partial wave expansion on the form factors $F_i$ using the associated Legendre polynomials:
\begin{align}
& F_{0, t}(q^{2}, k^{2}, q \cdot \bar{k})=\sum_{l=0}^{\infty} \sqrt{2 l+1} F_{0, t}^{(l)}(q^{2}, k^{2}) P_{l}^{(0)}(\cos \theta_{\pi}),  \nonumber\\
& F_{\|}(q^{2}, k^{2}, q \cdot \bar{k})=\sum_{l=1}^{\infty} \sqrt{2 l+1} F_{\|}^{(l)}(q^{2}, k^{2}) \frac{P_{l}^{(1)}(\cos \theta_{\pi})}{\sin \theta_{\pi}}.\label{PartialWFF}
\end{align}
When this expansion is inserted into Eq.~(\ref{CorrFuncIm}), the two-body phase space integration
$\int d\tau_2$ will act as an S-wave projector so that the contribution of $F_{\|}$ vanishes. Applying
Eq.~(\ref{Fudpieta}), Eq.~(\ref{BtopietaFF}) and Eq.~(\ref{PartialWFF}) to Eq.~(\ref{CorrFuncIm}), and
performing the phase space integration, one arrives at
\begin{equation}
{\rm Im}\Pi_{\mu}^{\rm H}(s,q^{2})=\frac{i}{16\pi}\frac{\sqrt{\lambda_{\pi\eta}(s)}}{s}B_0 F^{*}_{\pi\eta}(s)\left[F_{t}^{(0)}(s,q^{2})\frac{q_{\mu}}{\sqrt{q^{2}}}+F_{0}^{(0)}(s,q^{2})\frac{2\sqrt{q^{2}}}{\sqrt{\lambda_{B}}}\left(p_{\mu}-\frac{p\cdot q}{q^{2}}q_{\mu}\right)\right].
\end{equation}
Using the dispersion relation, Eq.~(\ref{disperRela}), we can obtain the correlation function at the hadron level.

\subsection{Quark-Gluon Level}

At quark-gluon level,  the correlation function in Eq.~(\ref{CorrFunc}) should be calculated in the deep
Euclidean region with $p^2, q^2\ll 0$ utilizing the operator-product-expansion (OPE). For the $\bar B^0$ decay
the OPE is performed in the heavy quark limit with the bottom quark field being translated to an effective field
in the heavy quark effective theory (HQET), $b(x)={\rm exp}(-im_b v\cdot x)b_v(x)$, where $v$ is the four-velocity
of the $\bar B^0$. The correlation function is expressed as a convolution of a perturbative kernel and the
$B$ meson light-cone distribution amplitudes (LCDAs). 

Since the internal light quark propagates in the background field of soft gluons,  in the calculation
of the perturbative kernel, one should use a light quark propagator of the form
 \cite{Balitsky:1987bk,Khodjamirian:1998ji}: 
\begin{align}
\left\langle 0 |T\{q_{i}(x) \bar{q}_{j}(0)\}| 0\right\rangle=&\ \int \frac{d^{4} k}{(2 \pi)^{4}} e^{-i k x}
\frac{i \delta_{i j}}{\not k-m_{q}}-i g \int \frac{d^{4} k}{(2 \pi)^{4}} e^{-i k x} \int_{0}^{1} d u\
G^{\mu \nu}_{ij}(u x)\nonumber\\
&\times\left[\frac{1}{2} \frac{\slashed k+m_{q}}{\left(m_{q}^{2}-k^{2}\right)^{2}} \sigma_{\mu \nu}
+\frac{1}{m_{q}^{2}-k^{2}} u x_{\mu} \gamma_{\nu}\right]~,
\label{lightqprop}
\end{align}
where $i,j$ are the color indices, $u$ is a dimensionless parameter, $G_{\mu\nu}=G_{\mu\nu}^a t^a$ is the
gluon field strength tensor. Here, the fixed-point gauge $x\cdot A(x)=0$ is used.
\begin{figure}
\begin{center}
\includegraphics[width=0.65\columnwidth]{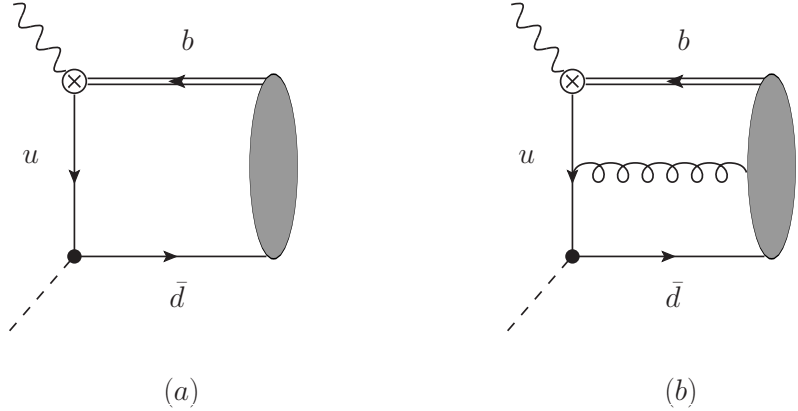} 
\caption{Diagrams for the calculation of the quark-gluon contribution to the two and three-body $B$ meson
  LCDAs. The circle cross denotes the $V-A$ current while the black dot denotes the interpolating current
  for the $\pi\eta$ state. The gray bubbles represent two- or three-particle LCDAs of the $B$ meson.}
\label{fig:BtopipiLCSR} 
\end{center}
\end{figure}
The corresponding diagrams are shown in Fig.\ref{fig:BtopipiLCSR}. In the diagram (a) the $u$ quark line is
a free propagator and the gray bubble denotes two-particle $B$ meson LCDAs. In the diagram (b) the $u$ quark
absorbs a soft gluon emitted from the $B$ meson state, and 
the gray bubble denotes the three-particle $B$ meson LCDAs.
The two and three-particle $B$ meson LCDAs are defined as \cite{Grozin:1996pq,Beneke:2000wa,Khodjamirian:2006st,Braun:2017liq}:
\begin{align}
&\left\langle 0\left|\bar{d}_{\alpha}(x)[x, 0] b_{v \beta}(0)\right| \bar{B}_{v}^{0}\right\rangle \nonumber\\
=&-\frac{i f_{B} m_{B}}{4} \int_{0}^{\infty} d \omega e^{-i \omega v \cdot x}\left\{(1+\slashed v)
\left(\phi_{+}^{B}(\omega)+\frac{\phi_{-}^{B}(\omega)-\phi_{+}^{B}(\omega)}{2 v \cdot x} \slashed x\right)
\gamma_{5}\right\}_{\beta \alpha},\nonumber\\
&\left\langle 0\left|\bar{d}_{\alpha}(x) G_{\lambda \rho}(u x) b_{v \beta}(0)\right| \bar{B}_v^{0}\right\rangle\nonumber\\
=&\ \frac{f_{B} m_{B}}{4} \int_{0}^{\infty} d \omega \int_{0}^{\infty} d \xi e^{-i(\omega+u \xi) v \cdot x}
\Big\{(1+\slashed v)\Big[\left(v_{\lambda} \gamma_{\rho}-v_{\rho} \gamma_{\lambda}\right)\left(\Psi_{A}(\omega, \xi)
-\Psi_{V}(\omega, \xi)\right)\nonumber\\
&-i \sigma_{\lambda \rho} \Psi_{V}(\omega, \xi)-\left(\frac{x_{\lambda} v_{\rho}-x_{\rho} v_{\lambda}}{v \cdot x}\right)
X_{A}(\omega, \xi)+\left(\frac{x_{\lambda} \gamma_{\rho}-x_{\rho} \gamma_{\lambda}}{v \cdot x}\right)
\left(Y_{A}(\omega, \xi)+W(\omega, \xi)\right)\nonumber\\
&-\frac{\slashed x}{(v\cdot x)^2}(x_{\lambda}v_{\rho}-x_{\rho}v_{\lambda}) W(\omega,\xi)+\frac{\slashed x}{(v\cdot x)^2}(x_{\lambda}\gamma_{\rho}-x_{\rho}\gamma_{\lambda}) Z(\omega,\xi)\nonumber\\
&+i\epsilon_{\lambda\rho\alpha\beta}\frac{x^{\alpha}}{v\cdot x}\left(v^{\beta}{\bar X}_A(\omega, \xi)-\gamma^{\beta}{\bar Y}_A(\omega, \xi)\right)\gamma_{5}\Big] \gamma_{5}\Big\}_{\beta \alpha},\label{BLCDAs}
\end{align}
where $\alpha,\beta$ are spinor indexes, $\omega$ and $\xi$ are the plus components of the light-quark and gluon
momentum, respectively, in the $B$ meson. $[x,0]$ is the Wilson line along the light-cone direction. As
illustrated in Ref.\cite{Colangelo:2000dp}, in the deep Euclidean region $p^2, q^2\ll 0$, the space-time interval
$x$ of the correlation function in Eq.~(\ref{CorrFunc})  is almost on the light-cone: $x^2\sim -1/q^2\to 0$.
Therefore, the fixed-point gauge $x\cdot A=0$  used for the light quark propagator in Eq.~(\ref{lightqprop})
is equivalent to the light-cone gauge $n\cdot A=0$, so that the Wilson line $[x,0]$ can be set to 1.
The explicit form of the $B$ meson LCDAs can be found in Appendix~A.

In Eq.~(\ref{BLCDAs}), the $v\cdot x$ in the denominator is difficult to be dealt with directly.
To overcome this difficulty, we define a new set of LCDAs as
\begin{equation}
{\psi}^{(1)}(\omega)=\int_{0}^{\omega}d\tau\ \psi(\tau),~~~~{\psi}^{(2)}(\omega)=\int_{0}^{\omega}d\tau\ \psi^{(1)}(\tau),~~~~\psi(\omega)=\frac{d^n}{d\omega^n}\psi^{(n)}(\omega)~,
\end{equation}
where $\psi$ can be any LCDA appears in Eq.~(\ref{BLCDAs}). The advantage of
this definition is that, with the help of integration by part, one can eliminate the $v\cdot x$ denominator:
\begin{equation}
\int_{0}^{\infty}d\omega\ e^{-i\omega v\cdot x}\frac{\psi(\omega)}{(v\cdot x)^{n}}\cdots=i^n\int_{0}^{\infty}d\omega\
e^{-i\omega v\cdot x}{\psi}^{(n)}(\omega)\cdots.
\end{equation}
The ellipsis denotes the terms independent of $\omega$. Note that during the integration by part,
since ${\psi}(0)=0$, the boundary term at $\omega=0$ vanishes trivially. On the other hand, it
seems that the boundary term at $\omega=\infty$ is nonzero since ${\psi}^{(n)}(\infty)$ is finite.
However, when $\omega\to \infty$, the exponential  ${\rm exp}(-i\omega v\cdot x)$ fluctuates heavily
during the integration of $x$. Thus the boundary term at $\omega=\infty$ is highly suppressed and
can be neglected.

Unlike for the $B$ meson, there is no systematic development of the $D$ meson LCDAs in the literature.
The only available one is the two-particle LCDA defined in Refs.~\cite{Zuo:2006re,Li:2008ts}:
\begin{align}
\langle0|\bar{u}_{\alpha}^{i}(x)c_{\beta}^{j}(0)|D^{0}(p)\rangle= \frac{i}{4N_{c}}f_{D}\delta_{ij}
\left[(\slashed p+m_{D})\gamma_{5}\right]_{\beta\alpha}\int_{0}^{1}du\ e^{-iup\cdot x}\varphi_{D}(u),
\end{align}
where $i,j$ are the color indices, and $f_D$ is the $D$ meson decay constant. $\varphi_D(u)$ is the
two-particle LCDA of $D$ meson, which has the form 
\begin{equation}
\varphi_D(u)=6u(1-u)\left[1-C_d(1-2u)\right]~.
\end{equation}
$C_d$ is a parameter ranging from 0 to 1, which can be fixed from experiments, $C_d=0.7$ \cite{Zuo:2006re}.

In the case of the $\bar B^0$ decay, the OPE calculation for the correlation function with contributions from
the two-particle LCDAs are
\begin{align}
&\Pi_{\mu}^{\rm OPE}(p,q)^{(2)} \nonumber\\
=&\  if_{B}m_{B}\int_{0}^{\infty}d\omega\frac{1}{\Delta_2}\left\{ \phi_{+}(\omega)\left[(m+\omega)v_{\mu}-p_{\mu}
\right]-4{\Phi^{(1)}}(\omega)v_{\mu}\right\} \nonumber \\
& -2if_{B}m_{B}\int_{0}^{\infty}d\omega\frac{1}{\Delta_2^{2}}(1-\frac{\omega}{m_{B}}){\Phi^{(1)}}(\omega)
\left[(m-\omega)p_{\mu}+\left(\frac{\omega}{m_B} (m_B^2+p^2-q^2)-p^{2}\right)v_{\mu}\right]\label{OPE2body}~,
\end{align}
where $m$ is the $u$ quark mass $m_u$ in the $\overline{MS}$ scheme at the scale $\mu=2$ GeV, which is
given by the Particle Data Group (PDG) \cite{Zyla:2020zbs}. For the $D^0$ decay $m$ denotes the $d$
quark mass $m_d$.
The discontinuity of the correlation function across the complex plane of $p^2$ comes from the denominators $1/\Delta_2^n$. On the other hand, the correlation function contributed by the three-particle LCDAs contains the similar denominators $1/\Delta_3^n$. However, it is much more involved so we do not show it here explicitly. These denominators read
\begin{align}
&\Delta_3 =\left[p-(\omega+u\xi) v\right]^{2}-m^{2}=\left(1-\frac{\omega+u\xi}{m_{B}}\right)\left[p^{2}-\Sigma(\omega,\xi,
q^{2},m^2)\right],~~\Delta_2=\Delta_3(\xi=0),\nonumber \\
&\Sigma(\omega,\xi, q^{2},m^2) =\frac{\omega+u\xi}{m_{B}-\omega-u\xi}(m_{B}^{2}-q^{2})
-\frac{m_{B}[(\omega+u\xi)^2-m^{2}]}{m_{B}-\omega-u\xi}~.
\end{align}

Extracting the discontinuity across the complex plane of $p^2$ for the correlation function is equivalent to extracting the discontinuity of $1/\Delta_{2,3}^n$. For the single power terms $1/\Delta_{2,3}$, their
discontinuity can be obtained by the replacement
\begin{equation}
\frac{1}{\Delta_3}\to (-2\pi i)\frac{m_B}{m_B-\omega-u\xi}\delta\left(p^{2}-\Sigma(\omega,\xi,q^{2},m^2)\right),
\end{equation}
and for $1/\Delta_{2}$ the replacement is the same but with $\xi=0$. To extract the discontinuity of
the higher power terms $1/\Delta_{2,3}^n$, we can firstly transform it into the form of a derivative on
the $1/\Delta_{2,3}$, which reads
\begin{equation}
{\rm Disc}\left[\int \frac{1}{\Delta_{3}^{n}}(\cdots)\right]=\frac{(-2\ensuremath{\pi}i)}{(n-1)!}
\left(\frac{\partial}{\partial\Omega}\right)^{n-1}\int \text{\ \ensuremath{\delta}}\left(
\left[p-(\omega+u\xi)v\right]^{2}-\Omega\right)(\cdots)\Big|_{\Omega=m^{2}}~,
\end{equation}
where $(\cdots)$ denotes all the terms except $1/\Delta_{2,3}^n$ in the integrand of Eq.~(\ref{OPE2body}). Finally, we obtain the correlation function at the quark-gluon level by the
dispersion relation
\begin{equation}
\Pi_{\mu}^{\rm OPE} (p,q)=\frac{1}{\pi}\int_{0}^{\infty}ds\ \frac{{\rm Im}\Pi_{\mu}^{\rm OPE} (s,q^{2})}{s-p^{2}-i\epsilon}~,
\label{disperRelaOPE}
\end{equation}
which should be equal to that given in Eq.~(\ref{disperRela}): $\Pi_{\mu}^{\rm H}=\Pi_{\mu}^{\rm OPE}$. According to
the quark-hadron duality,  the dispersion integral in the region $s_0^{\pi\eta}<s<\infty$ on the two sides
are canceled, where $s_0^{\pi\eta}=4 m_K^2$ is the threshold of $K\bar K$, which is the lowest two-meson
state above the $\pi\eta$ state.  After Borel transformation, one arrives at the sum rules equation. For $F_{0}^{(0)}(s,q^{2})$ it reads as
\begin{align}
 & \frac{1}{16\pi^{2}}\int_{(m_{\pi}+m_{\eta})^{2}}^{s_{0}^{\pi\eta}}ds\ e^{-s/M^{2}}\frac{\sqrt{\lambda_{\pi\eta}(s)}}{s}B_{0}F_{\pi\eta}^{*}(s)F_{0}^{(0)}(s,q^{2})\frac{2\sqrt{q^{2}}}{\sqrt{\lambda_{B}}}\nonumber \\
= &\  f_{B}\sum_{n=0}^{1}\left(m_B^{2}\frac{\partial}{\partial\Omega}\right)^n\int_{0}^{\infty}d\omega\ \Theta\left(\omega,0,s_{0}^{\pi\eta},M^2,q^2,m^2\right)I_{2}^{(n)}(\omega,q^{2})\Big|_{\Omega=m^{2}}\nonumber \\
 & +f_{B}\sum_{n=1}^{3}\frac{1}{n!}\left(m_B^{2}\frac{\partial}{\partial\Omega}\right)^n \int_{0}^{\infty}d\omega d\xi\ \Theta\left(\omega,\xi,s_{0}^{\pi\eta},M^2,q^2,\Omega\right)I_{3}^{(n)}(\omega,\xi,q^{2},\Omega) \Big|_{\Omega=m^{2}}~, \label{F0match}
\end{align}
where 
\begin{equation}
\Theta\left(\omega,\xi,s_{0}^{\pi\eta},M^2,q^2,\Omega\right)=e^{-\Sigma(\omega,\xi,q^{2},\Omega)/M^{2}}\theta\left(\Sigma(\omega,\xi,q^{2},\Omega)\right)\theta\left(s_{0}^{\pi\eta}-\Sigma(\omega,\xi,q^{2},\Omega)\right).\label{thetaFunc}
\end{equation}
For $F_{t}^{(0)}(s,q^{2})$, the sum rules equation is similar, it reads
\begin{align}
 & \frac{1}{16\pi^{2}}\int_{(m_{\pi}+m_{\eta})^{2}}^{s_{0}^{\pi\eta}}ds\ e^{-s/M^{2}}\frac{\sqrt{\lambda_{\pi\eta}(s)}}{s}B_{0}F_{\pi\eta}^{*}(s)\left(F_{t}^{(0)}(s,q^{2})\frac{1}{\sqrt{q^{2}}}-F_{0}^{(0)}(s,q^{2})\frac{2p\cdot q}{\sqrt{\lambda_{B}q^{2}}}\right)\nonumber \\
= &\  f_{B}\sum_{n=0}^{1}\left(m_B^{2}\frac{\partial}{\partial\Omega}\right)^n\int_{0}^{\infty}d\omega\ \Theta\left(\omega,0,s_{0}^{\pi\eta},M^2,q^2,m^2\right)J_{2}^{(n)}(\omega,q^{2})\Big|_{\Omega=m^{2}}\nonumber \\
 & +f_{B}\sum_{n=1}^{3}\frac{1}{n!}\left(m_B^{2}\frac{\partial}{\partial\Omega}\right)^n \int_{0}^{\infty}d\omega d\xi\ \Theta\left(\omega,\xi,s_{0}^{\pi\eta},M^2,q^2,\Omega\right)J_{3}^{(n)}(\omega,\xi,q^{2},\Omega) \Big|_{\Omega=m^{2}}~. \label{F0match}
\end{align}
The explicit expression for the $I_{2,3}^{(n)}$ and $J_{2,3}^{(n)}$ functions are given in  Appendix~B. 

On the other hand, for the $D^0$ decay the treatment is more simple because there is only one
two-particle LCDA $\varphi_D$ and no $v\cdot x$ terms appear in the denominator.  The sum rule
equations for the $D^0$ decay are
\begin{align}
 & \frac{1}{16\pi^{2}}\int_{(m_{\pi}+m_{\eta})^{2}}^{s_{0}^{\pi\eta}}ds\ e^{-s/M^{2}}\frac{\sqrt{\lambda_{\pi\eta}(s)}}{s}B_{0}F_{\pi\eta}^{*}(s)F_{0}^{(0)}(s,q^{2})\frac{2\sqrt{q^{2}}}{\sqrt{\lambda_{B}}}\nonumber \\
= &\  f_{D}\int_{0}^{1}\frac{du}{\bar u}\ \Theta^{\prime}\left(u,s_{0}^{\pi\eta},M^2,q^2,m^2\right)(m-\bar u m_D)\varphi_D(u)~,\nonumber \\
& \frac{1}{16\pi^{2}}\int_{(m_{\pi}+m_{\eta})^{2}}^{s_{0}^{\pi\eta}}ds\ e^{-s/M^{2}}\frac{\sqrt{\lambda_{\pi\eta}(s)}}{s}B_{0}F_{\pi\eta}^{*}(s)\left(F_{t}^{(0)}(s,q^{2})\frac{1}{\sqrt{q^{2}}}-F_{0}^{(0)}(s,q^{2})\frac{2p\cdot q}{\sqrt{\lambda_{B}q^{2}}}\right)\nonumber \\
= &\  f_{D}\int_{0}^{1}\frac{du}{\bar u}\ \Theta^{\prime}\left(u,s_{0}^{\pi\eta},M^2,q^2,m^2\right)(m+u m_D)\varphi_D(u)~.\label{DF0tmatch}
\end{align}
where $\bar u=1-u$ and
\begin{align}
&\Theta^{\prime}\left(u,s_{0}^{\pi\eta},M^{2},q^{2},m^{2}\right)\nonumber\\
=&\  e^{-\left(um_{D}^{2}+\frac{m^{2}-uq^{2}}{\bar{u}}\right)/M^{2}}\theta\left(um_{D}^{2}+\frac{m^{2}-uq^{2}}{\bar{u}}\right)\theta\left(s_{0}^{\pi\eta}-um_{D}^2-\frac{m^{2}-uq^{2}}{\bar{u}}\right).\label{DthetaFunc}
\end{align}

From the sum rules Eqs.~(\ref{F0match}) and (\ref{DF0tmatch}), it can be concluded that the convolution on the left hand side must be real. This means that the strong phase of the transition form factor
and the $\pi\eta$ scalar form factor must cancel each other:
\begin{equation}
{\rm Im}\left[F_{\pi\eta}^{*}(s)F_{0,t}^{(0)}(s,q^{2})\right]=0~.
\label{cancelphase}
\end{equation}
This result is exactly the one deduced from the Watson--Migdal theorem~\cite{Watson:1952ji,Migdal:1956tc}.
We thus can conclude that since the $\pi\eta$ system is decoupled from the leptons, the phase
measured in $\pi\eta$ elastic scattering and the transition matrix element $B^0\to \pi\eta$ must be equal.

\subsection{Comparison with the LCSR in the Narrow-Width Limit}

In the narrow-width limit, instead of the $1\to 2$ process as given in Eq.~(\ref{BtopietaFF}), one
only considers the $1\to 1$ process $\bar B^0\to a_0(980)$ which is parameterized as
\begin{align}
-i\left\langle a_0(p)|\bar{u} \gamma_{\mu} \gamma_{5} b | \bar{B}^{0}(q+p)\right\rangle= F_{+}(q^2) p_{\mu}+F_{-}(q^2) q_{\mu}.\label{BtopietaFFnarrowWidth}
\end{align}
Using this parameterization, one obtains the sum rules equations. For example, the sum rule
for $F_{+}(q^2)$ is
\begin{equation}
m_{a_{0}}f_{a_{0}}e^{-m_{a_{0}}^{2}/M^{2}}F_{+}(q^{2})=\frac{1}{\pi}\int_{0}^{s_{0}^{\pi\eta}}ds\ e^{-s/M^2}
{\rm Im}\Pi_{\mu}^{\rm OPE}(s,q^{2})~.
\label{FplusNarrSR}
\end{equation}

The sum rule Eq.~(\ref{F0match}) must be consistent with the above one in the narrow-width limit. To realize
this matching, one should find a suitable parameterization for $F_{0,t}^{(0)}(s,q^{2})$, which must have an
explicit dependence on the $a_0$ decay width $\Gamma_{a_0}^{\rm tot}$ so that it has a definite limit when
$\Gamma_{a_0}^{tot}\to0$.  Following the method from Refs.~\cite{Cheng:2017smj,Cheng:2019tgh} and with the
constraint of Eq.~(\ref{cancelphase}), instead of directly using the numerical result in Fig.~\ref{fig:pietaFF}
one can temporarily assume the form of  $F_{\pi\eta}^{\bar{d}u}(s)^{*}$ as 
\begin{equation}
F_{\pi\eta}^{*}(s)=\frac{g_{a_{0}\pi\eta}m_{a_{0}}f_{a_{0}}}{m_{a_{0}}^{2}-s+i\sqrt{s}\Gamma_{a_{0}}(s)}e^{-i\phi_{a_{0}}(s)},\label{ParFpieta}
\end{equation}
where $g_{a_{0}\pi\eta}$ is the strong coupling of the $a_0$ to $\pi\eta$, and $f_{a_{0}}$ is the decay constant of the $a_0$:
\begin{equation}
\langle0|\bar{d}u|a_{0}\rangle=m_{a_{0}}f_{a_{0}},
\end{equation}
and 
\begin{equation}
\Gamma_{a_{0}}(s)=\frac{g_{a_{0}\pi\eta}^{2}\sqrt{\lambda_{\pi\eta}(s)}}{16\pi s^{3/2}}\theta\left(s-(m_{\pi}+m_{\eta})^{2}\right)\equiv\Gamma_{a_{0}}^{tot}\frac{m_{a_0}^3}{s^{3/2}}\sqrt{\frac{\lambda_{\pi\eta}(s)}{\lambda_{\pi\eta}(m_{a_0}^2)}}\theta\left(s-(m_{\pi}+m_{\eta})^{2}\right)
\end{equation}
is the decay width of the $a_0\to \pi\eta$ with the mass squared of the $a_0$ being a variable $s$.
Accordingly, to satisfy Eq.~(\ref{cancelphase}), $F_0^{(0)}(s,q^2)$ must have an inverse complex phase,
which reads 
\begin{equation}
F_{0}^{(0)}(s,q^{2})\frac{2\sqrt{q^{2}}}{\sqrt{\lambda_{B}}}
=\frac{g_{a_{0}\pi\eta}F_{+}(q^{2})}{m_{a_{0}}^{2}-s-i\sqrt{s}\Gamma_{a_{0}}(s)}e^{i\phi_{a_{0}}(s)}~.
\label{ParF00}
\end{equation}
Thus using Eq.~(\ref{cancelphase}) we can obtain the phase of $F_{0,t}^{(0)}(s,q^{2})$ as
\begin{equation}
\phi_{a_0}(s)=\delta_{\pi\eta}(s)-{\rm arctan}\left[\frac{{\sqrt s}\ \Gamma_{a_0}(s)}{m_{a_0}^2-s}\right]~,
\label{phaseofFF}
\end{equation}
where $\delta_{\pi\eta}(s)$ is the phase of $F_{\pi\eta}(s)$, which is taken from Fig.~\ref{fig:pietaFF}.
After inserting Eqs.~(\ref{ParFpieta}) and (\ref{ParF00}) into Eq.~(\ref{F0match}), and taking the
narrow-width limit of $\Gamma_{a_0}^{\rm tot}\to0$, one arrives at exactly the same sum rule as
Eq.~(\ref{FplusNarrSR}). It should be mentioned that for Eq.~(\ref{ParFpieta}) one should in principle
multiply it by an additional factor on the right-hand-side to make the
normalization consistent with that given in  Fig.~\ref{fig:pietaFF}. However, this factor can be absorbed
by $F_{\pm}(q^2)$ in Eq.~(\ref{ParF00}) so we neglect it here. Similarly, the $F_t^{(0)}(s,q^2)$ can
be parameterized by $F_-(q^2)$ as
\begin{equation}
F_{t}^{(0)}(s,q^{2})\frac{1}{\sqrt{q^{2}}}-F_{0}^{(0)}(s,q^{2})\frac{2p\cdot q}{\sqrt{\lambda_{B}q^{2}}}
=\frac{g_{a_{0}\pi\eta}F_{-}(q^{2})}{m_{a_{0}}^{2}-s-i\sqrt{s}\Gamma_{a_{0}}(s)}e^{i\phi_{a_{0}}(s)}~.
\label{ParF0t}
\end{equation}
Finally, inserting Eqs.~(\ref{ParF00}) and~(\ref{ParF0t}) into the sum rules equations: Eqs.~(\ref{F0match}) and the one for three-body LCDA contribution, one can extract the $F_{\pm}(q^{2})$.  

Note that  since the form factors obtained here are only applicable to the small $q^2$ region due to the
light-cone OPE, one should extent their applicability to the physical region with larger $q^2$.
In Eq.~(\ref{ParF00}) the dependence on $s$ and $q^2$ are factorized, which enables us to find an
appropriate parameterization for $F_{\pm}(q^{2})$ to realize the extension of $q^2$. This factorization
approach can also be used for the $D^0$ decay.

\section{Numerical Results}

\begin{figure}
\begin{center}
\includegraphics[width=0.6\columnwidth]{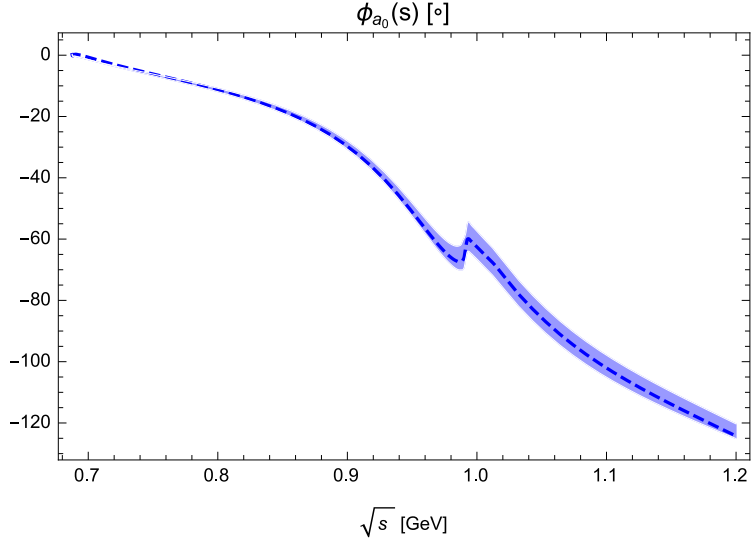} 
\caption{The phase of the $B^0/D^0\to \pi\eta$ form factors, with the blue band showing the uncertainty from
  the Borel parameter and the LECs of ChPT.}
\label{fig:phaseFF} 
\end{center}
\end{figure}

The parameters used are: $m_{\pi}=0.139$~GeV, $m_K=0.496$~GeV, $m_{a_0}=0.98$~GeV, $m_B=5.28$~GeV, $m_D=1.87$~GeV,
$m_u=2.2$~MeV, $m_d=4.7$~MeV  \cite{Zyla:2020zbs};
$f_B=0.207$~GeV \cite{Gelhausen:2013wia}, $f_D=0.207$~GeV \cite{Zuo:2006re}; $g_{{a_0}\pi\eta}
=m_{a_0}{\sqrt{8\pi {\bar g_{\eta}}}}=3.307$~GeV \cite{Baru:2004xg} with $\bar g_{\eta}=0.453$ \cite{Achasov:2002ir};
$B_0=1.7$~GeV \cite{Colangelo:2000dp}. According to Eq.~(\ref{phaseofFF}), we can obtain the phase of the form factors.
The numerical
result for $\phi_{a_0}(s)$ is shown in Fig.~\ref{fig:phaseFF}, which is nonzero above the $\pi\eta$ threshold.
The blue band shows the uncertainty from the Borel parameter and the LECs of ChPT.

In the calculation of $F_{\pm}(q^2)$, instead of infinity the upper limits of the $\omega$ and $\xi$
integrations are chosen as $\omega+\xi<2\omega_0$ with $\omega_0=(2/3)\bar \Lambda$
\cite{Lu:2018cfc}, where $\bar \Lambda=m_B-m_b=0.45$~GeV in the heavy quark limit \cite{Ball:1993xv}.
It can be found that numerically the different integration upper limits $2\omega_0$ and $\infty$ lead
to almost the same result. The reason is that at large energies, the original LCDAs: $\phi_+$, $\psi_{V,A}$  are highly suppressed by their exponential behavior, while the contributions from the newly defined LCDAs such as $\Phi^{(1)}$, $X_A^{(1)}$, $Y_A^{(1)}$... are cut off by the theta functions of Eq.~(\ref{thetaFunc}).  The Borel parameter $M$ is taken so that the fraction of the pole
contribution is around 40$\%$: 
\begin{equation}
\int_{0}^{s_{th}^{\pi\eta}}ds\ e^{-s/M^{2}}{\rm Im}\Pi_{\mu}^{\rm OPE}(s,q^{2})\Big/\int_{0}^{\infty}ds\
e^{-s/M^{2}}{\rm Im}\Pi_{\mu}^{\rm OPE}(s,q^{2})\approx40\%~.
\end{equation}
Note that this fraction is an empirical value, practically one should allow a finite region for the choice of
the Borel parameter.  Accordingly, we choose the region of the Borel parameters as $1.0~{\rm GeV}<M<1.2~{\rm GeV}$
for the $\bar B^0$ decay and $1.2~{\rm GeV}<M<1.4~{\rm GeV}$ for the $D^0$ decay, where the center value of each
region corresponds to the empirical fraction. The numerical result will depend on the Borel parameter, and the
regions for the Borel parameters are used to estimate the error of the numerical results.

\begin{figure}
\begin{center}
\includegraphics[width=0.45\columnwidth]{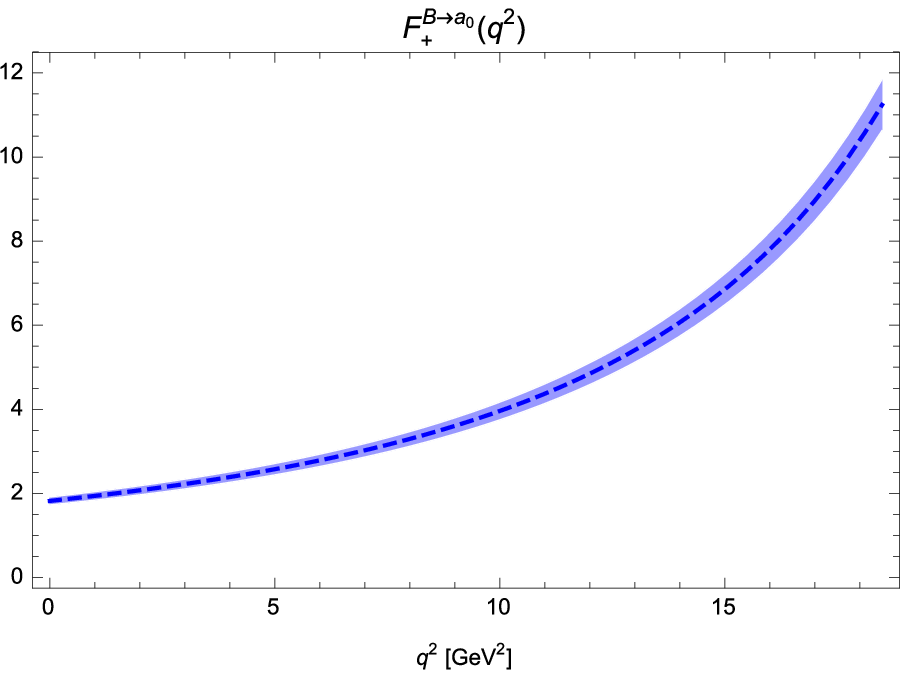} 
\includegraphics[width=0.45\columnwidth]{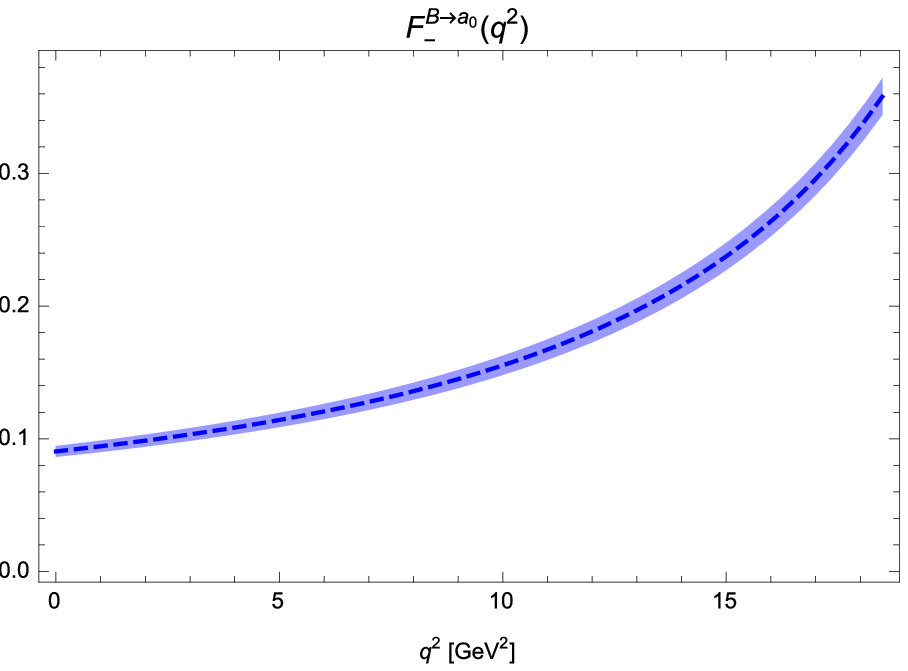} 
\caption{$F_+(q^2)$ (left) and $F_-(q^2)$ (right) for the $\bar B^0$ decay. The blue bands reflect the
  uncertainty from the Borel parameter and the LECs of ChPT.}
\label{fig:FFpmB} 
\end{center}
\end{figure}

\begin{figure}
\begin{center}
\includegraphics[width=0.44\columnwidth]{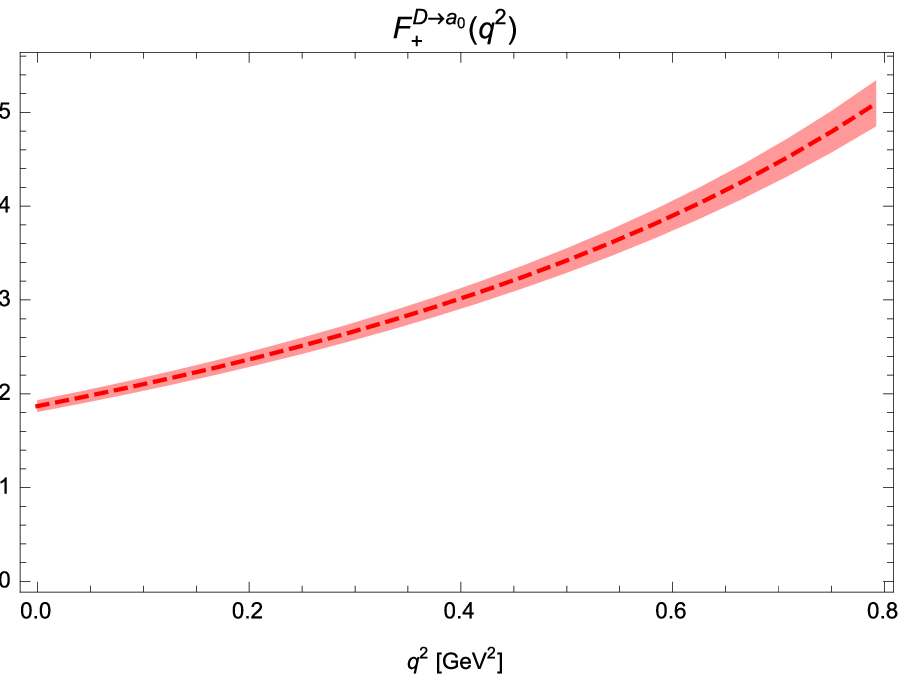} 
\includegraphics[width=0.46\columnwidth]{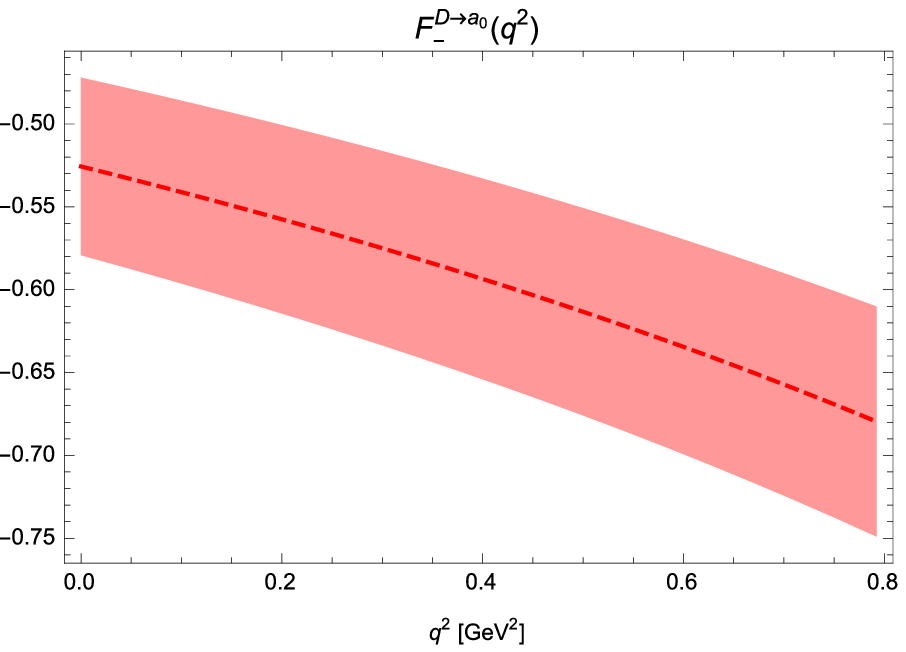} 
\caption{$F_+(q^2)$ (left) and $F_-(q^2)$ (right) for the $D^0$ decay. The red bands reflect the uncertainty
  from the Borel parameter and the LECs of ChPT.}
\label{fig:FFpmD} 
\end{center}
\end{figure}

To extend the applicability of $F_{\pm}(q^2)$ to the whole physical region for $q^2$, one should use a suitable
parameterization for $F_{\pm}(q^2)$ to realize the extension. One of the popular parameterization is
the $z-$series expansion \cite{Bourrely:2008za}:
\begin{equation}
  F_{\pm}(q^{2})=\frac{F_{\pm}(0)}{1-\frac{q^{2}}{m_{\rm fit}^{2}}}\left[1+b_{\pm} \zeta\left(q^{2}\right)
    +c_{\pm} \zeta^{2}\left(q^{2}\right)\right], \label{fitParfunc}
\end{equation}
where
\begin{align}
\zeta\left(q^{2}\right)=z\left(q^{2}\right)-z(0),~~~
z\left(q^{2}\right)=\frac{\sqrt{t_{+}-q^{2}}-\sqrt{t_{+}-t_{0}}}{\sqrt{t_{+}-q^{2}}+\sqrt{t_{+}-t_{0}}}
\end{align}
with $t_{\pm} \equiv(m_{B_{s}} \pm m_{f_{0}})^{2}$ and $t_0=t_+(1-\sqrt{1-t_-/t_+})$.  The fitting regions are
chosen as $0~{\rm GeV^2}<q^2<2~{\rm GeV^2}$ for  $F_{\pm}^{\bar B^0\to a_0}$,  $0~{\rm GeV^2}<q^2<0.7~{\rm GeV^2}$
for $F_{+}^{D^0\to a_0}$  and $0\  {\rm GeV}<q^2<0.3\   {\rm GeV}$ for $F_{-}^{D^0\to a_0}$.  The fit
results are listed in Table~\ref{tab:fitFF}, and the corresponding curves are shown in Fig.~\ref{fig:FFpmB}
and Fig.~\ref{fig:FFpmD}. 
\begin{table}
 \caption{Fit results for the $F_{\pm}(q^2)$}\label{tab:fitFF}
\begin{tabular}{|c|cccc|}
\hline 
\hline 
Form Factors & $F_{\pm}(0)$ & $m_{\rm fit}$ & $b_{\pm}$ & $c_{\pm}$\tabularnewline
\hline 
$F^{\bar B^0\to a_0}_{+}$ & $1.82\pm0.1$ & $m_B$ & $-4.18\pm0.02$ & $16.29\pm0.04$\tabularnewline
$F^{\bar B^0\to a_0}_{-}$ & $0.09\pm0.004$ & $m_B$ & $-0.77\pm0.04$ & $8.22\pm0.8$\tabularnewline
\hline 
$F^{D^0\to a_0}_{+}$ & $1.87\pm0.07$ & $1.33\pm 0.03$ & $-19.17\pm 0.8$ & $0$\tabularnewline
$F^{D^0\to a_0}_{-}$ & $-0.53\pm0.06$ & $m_D$ & $0$ & $0$\tabularnewline
\hline 
\end{tabular}
\end{table}
Note that in Ref.~\cite{Bourrely:2008za} $m_{\rm fit}$ is fixed as the pole mass,
which is equal to $m_B$ or $m_D$ for the $B$ or $D$ decays, respectively. Here, we try to choose it as a
free parameter for the fit to improve the fit result.  However, since the phase space of $q^2$ for the $B$
decay is very large, the fitted mass pole $m_{\rm fit}^2$ may be smaller than the upper limit of $q^2$,
so that it will cause a singularity during the phase space integration. 
Thus we just fix it as $m_B$ in the fitting.  On the other hand, we find that for the $D^0$ form factors,
the fitted $b_-$
and $c_{\pm}$ are extremely large.  A possible reason is that the $D^0$ form factors are insensitive to the
terms proportional to $b_-\zeta\left(q^{2}\right)$ and $c_{\pm}\zeta^2\left(q^{2}\right)$ in Eq.~(\ref{fitParfunc}).
Therefore, for the fit of $F_{+}^{D^0\to a_0}$, we fix $c_{\pm}=0$, while for the fit
of $F_{-}^{D^0\to a_0}$, we fix $b_{\pm}=c_{\pm}=0$ and $m_{\rm fit}=m_D$.

Finally, we calculate the decay branching fraction of  $\bar B^0/D^0 \to (a_0(980)\to\pi^{\pm}\eta)l\bar\nu$.
The expression of the differential decay width reads:
\begin{align}
\frac{d^{3}\Gamma(\bar B^0/D^0 \to \pi^{\pm}\eta\  l^{\mp} \nu)}{dq^{2}dk^{2}d{\rm cos}\theta_{\pi}}=&\frac{2G_{F}^{2}|V_{bu/cd}|^{2}(q^{2}-m_{l}^{2})^{2}\sqrt{\lambda_{B/D}\lambda_{\pi\eta}}}{32(2\pi)^{5}m_{B/D}^{3}k^{2}q^{2}}\nonumber\\
&\times\left[F_{0}^{0}(k^2,q^{2})^{2}{\rm sin^{2}}\theta_{\pi}+\frac{m_{l}^{2}}{q^{2}}\left(F_{t}^{0}(k^2,q^{2})-F_{0}^{0}(k^2,q^{2}){\rm cos}\theta_{\pi}\right)^2\right]~,
\end{align}
where $m_l$ is the lepton mass and it is chosen as zero for $e$ and $\mu$. The numerical result of the
total decay widths are
\begin{align}
{\cal B}(D^0 \to \pi^{-}\eta\  l^{+} \nu)&=(1.36\pm0.21)\times 10^{-4},\nonumber\\
{\cal B}(\bar B^0 \to \pi^{+}\eta\  l^{-} \nu)&=(0.60\pm0.07)\times 10^{-4}.
\end{align}
We note that the branching fraction obtained above for the $D^0$ decay is consistent with that
measured by the BESIII Collaboration \cite{Ablikim:2018ffp}:
\begin{align}
{\cal B}(D^0 \to \pi^{-}\eta\  e^{+} \nu)=(1.33\pm0.09)\times 10^{-4}.
\end{align}
This agreement is quite amazing. It is known that the $D$ meson LCDA are not developed as well as those
of the $B$ meson. One may expect that the result of $D^0$ decay contains larger uncertainties.
However, although the error of the $D^0$ decay branching fraction is one order larger than that of the
$\bar B^0$ decay, it is still small enough to give consistency  with the
experimental result. Furthermore,  we find that in the numerical calculation for the $B$ decay, the contribution
from the 3-particle LCDA is two orders of magnitude smaller than the one from the 2-particle LCDA. Therefore we
also do not expect a sizeable contribution from such corrections in the $D$ meson case.
On the other hand, up to now there is no experimental measurement on the branching fraction for
$\bar B^0 \to (a_0(980)\to\pi^{+}\eta)  e^{-} \nu$, so our result can be tested by future experiments.

\section{Conclusions}
In this work, we have analyzed the semi-leptonic decays $\bar B^0/D^0 \to (a_0(980)^{\pm}\to\pi^{\pm}\eta)
l^{\mp} \nu$ within light-cone sum rules. In the calculation at the quark-gluon level, we used the two-
and three-body LCDAs of the $B$ meson for the $\bar B^0$ decay and the only available two-body LCDA of the
$D$ meson for the $D^0$ decay. In the calculation at the hadron level, to include the finite-width effect
of the $a_0(980)$, we use a scalar form factor to describe the final-state interaction between in the
$\pi\eta$ system, which was previously calculated in the framework of unitarized Chiral Perturbation Theory.
The  resulting value of the decay branching fraction of the $D^0$ decay shows an amazing agreement with the one
measured by the BESIII Collaboration, while the branching fraction of the $\bar B^0$ decay can be tested
in future experiments.

\label{sec:conclusions}

\section*{Acknowledgements}
We are very grateful to Prof. Wei Wang, Prof. Zhen-Xing Zhao and Dr. Chien-Yeah Seng
for useful discussions. We are also grateful to Prof. Yu-Ming Wang and Dr. Yao Ji for introducing us
to the latest version of the $B$ meson LCDAs.  This work is supported in part by the NSFC and the
Deutsche Forschungsgemeinschaft
(DFG, German Research Foundation) through the funds provided to the Sino-German Collaborative  
Research Center TRR~110 “Symmetries and the Emergence of Structure in QCD”
(NSFC Grant No. 12070131001, DFG Project-ID 196253076 - TRR 110),  by the Chinese Academy of Sciences
(CAS) through a President's International Fellowship Initiative (PIFI) (Grant No. 2018DM0034), by the
VolkswagenStiftung (Grant No. 93562), and by the EU Horizon 2020 research and innovation programme,
STRONG-2020 project under grant agreement No. 824093.

\begin{appendix}
	
\section{LCDAs of $B$ and $D$ mesons\label{sec:LCDAofBD}}
In this appendix we list the explicit expressions for the LCDAs  of the $B$ meson \cite{Braun:2017liq}. The two-particle LCDAs are
\begin{eqnarray}
\phi_B^{+}(\omega) &=& {\omega \over \omega_0^2} \,  e^{-\omega/\omega_0} \,, \nonumber \\
\phi_B^{-}(\omega) &=& {1 \over \omega_0} \,  e^{-\omega/\omega_0}
- {\lambda_E^2 - \lambda_H^2 \over 9 \, \omega_0^3}  \,
\left [ 1 - 2 \, \left ( {\omega \over \omega_0} \right )
+ {1 \over 2} \, \left ( {\omega \over \omega_0} \right )^2 \right ]
\,  e^{-\omega/\omega_0}~,
\end{eqnarray}
where $2 \, \bar \Lambda^2 = 2 \, \lambda_E^2 + \lambda_H^2$ and $ \lambda_H^2=2 \lambda_E^2$. 
The three-particle LCDAs are classified by different twists:
\begin{eqnarray}
\Phi_3(\omega, \xi) &=&  \Psi_A(\omega, \xi) - \Psi_V(\omega, \xi) \,, \nonumber \\
\Phi_4(\omega, \xi) &=&  \Psi_A(\omega, \xi) + \Psi_V(\omega, \xi) \,, \nonumber \\
\Psi_4(\omega, \xi) &=&  \Psi_A(\omega, \xi) + X_A(\omega, \xi) \,, \nonumber \\
\bar{\Psi}_4(\omega, \xi) &=&  \Psi_V(\omega, \xi) - \bar{X}_A(\omega, \xi) \,, \nonumber \\
\Phi_5(\omega, \xi) &=&  \Psi_A(\omega, \xi) + \Psi_V(\omega, \xi)
+ 2 \, \left  [ Y_A -  \bar{Y}_A +  W \right ] (\omega, \xi)\,, \nonumber \\
\Psi_5(\omega, \xi) &=&  - \Psi_A(\omega, \xi) + X_A(\omega, \xi)
- 2 \, Y_A(\omega, \xi) \,, \nonumber \\
\bar{\Psi}_5(\omega, \xi) &=&  - \Psi_V(\omega, \xi) - \bar{X}_A(\omega, \xi)
+ 2 \, \bar{Y}_A(\omega, \xi) \,, \nonumber \\
\Phi_6(\omega, \xi) &=&  \Psi_A(\omega, \xi) - \Psi_V(\omega, \xi)
+ 2 \, \left  [ Y_A  +  \bar{Y}_A
+  W - 2 \, Z \right ] (\omega, \xi)  \,.
\label{3P B-meson DAs of definite twist}
\end{eqnarray}
Each LCDA with definite twist has the explicit form of:
\begin{eqnarray}
\Phi_3(\omega, \xi) &=&
{\lambda_E^2 - \lambda_H^2 \over 6 \, \omega_0^5} \, \omega \, \xi^2 \,
e^{-(\omega + \xi)/\omega_0} \,, \nonumber \\
\Phi_4(\omega, \xi) &=&
{\lambda_E^2 + \lambda_H^2 \over 6 \, \omega_0^4} \, \xi^2 \,
e^{-(\omega + \xi)/\omega_0} \,, \nonumber \\
\Psi_4(\omega, \xi) &=&
{\lambda_E^2 \over 3 \, \omega_0^4} \, \omega \, \xi \,
e^{-(\omega + \xi)/\omega_0} \,, \nonumber \\
\bar{\Psi}_4(\omega, \xi) &=&
{\lambda_H^2 \over 3 \, \omega_0^4} \, \omega \, \xi \,
e^{-(\omega + \xi)/\omega_0} \,, \nonumber \\
\Phi_5(\omega, \xi)
&=& {\lambda_E^2 + \lambda_H^2 \over 3 \, \omega_0^3} \, \omega \,
e^{-(\omega + \xi)/\omega_0} \,, \nonumber \\
\Psi_5(\omega, \xi)
&=& - {\lambda_E^2 \over 3 \, \omega_0^3} \, \xi \,
e^{-(\omega + \xi)/\omega_0} \,, \nonumber \\
\bar{\Psi}_5(\omega, \xi)
&=& - {\lambda_H^2 \over 3 \, \omega_0^3} \, \xi \,
e^{-(\omega + \xi)/\omega_0} \,, \nonumber \\
\Phi_6(\omega, \xi)
&=& {\lambda_E^2 - \lambda_H^2 \over 3 \, \omega_0^2} \,
e^{-(\omega + \xi)/\omega_0} \,.
\end{eqnarray}

\section{$I,J$ functions in the OPE calculation\label{sec:IJfunc}}
The explicit expressions for the $I,J$ functions in the OPE calculation are 
\begin{align}
& I_{2}^{(0)}(\omega) =4{\Phi}^{(1)}(\omega)-\phi_{+}(\omega)(m-m_{B}+\omega),\nonumber \\
& J_{2}^{(0)}(\omega) =4{\Phi}^{(1)}(\omega)-\phi_{+}(\omega)(m+\omega),\nonumber \\
& I_{2}^{(1)}(\omega,q^{2},\Omega) =2{\Phi}^{(1)}(\omega)\left(1-\frac{\omega}{m_{B}}\right)\left[\frac{m}{m_{B}}-\frac{\Sigma_0}{m_{B}^{2}}+\frac{\omega(\Sigma_0-q^{2})}{m_{B}^{3}}\right],\nonumber \\
& J_{2}^{(1)}(\omega,q^{2},\Omega) =2{\Phi}^{(1)}(\omega)\left(1-\frac{\omega}{m_{B}}\right)\left[\frac{\omega}{m_{B}}-\frac{\Sigma_0}{m_{B}^{2}}+\frac{\omega(\Sigma_0-q^{2})}{m_{B}^{3}}\right],\nonumber \\
&I_{3}^{(1)}(\omega,\xi,q^{2},\Omega)\nonumber \\ =&\ \frac{1}{m_B^4}\left[m_B^2\left(3(m(\Psi_A+\Psi_V)-{\bar X}_A^{(1)}-X_A^{(1)}+4{\bar Y}_A^{(1)}+4Y_A^{(1)})\right.\right.\nonumber \\
&\left.\left.+12(1-2u)W^{(1)}+(4u-3)\omega(\Psi_A-\Psi_V)+u(4u-3)\xi(\Psi_A-\Psi_V)+6u X_A^{(1)}-24u Y_A^{(1)}\right)\right.\nonumber \\
&\left.+m_B^3\left((1-4u)\Psi_A-(1+2u)\Psi_V)\right)+2m_B\Big((u-1)(q^2-\Sigma_{\xi})(\Psi_A-\Psi_V)\right.\nonumber\\
&\left.+3u(u\xi+\omega)\xi\left(4W^{(1)}-X_A^{(1)}+4Y_A^{(1)}\right)\Big)-2u(\omega+u\xi)(q^2-\Sigma_{\xi})(\Psi_A-\Psi_V)\right],\nonumber \\
&J_{3}^{(1)}(\omega,\xi,q^{2},\Omega)\nonumber \\ =&\ \frac{1}{m_B^4}\left[m_B^2\left(3(m(\Psi_A+\Psi_V)-{\bar X}_A^{(1)}-X_A^{(1)}+4{\bar Y}_A^{(1)}+4Y_A^{(1)})\right.\right.\nonumber \\
&\left.\left.+12(1-2u)W^{(1)}+(2u-3)\omega(\Psi_A-\Psi_V)+u(2u-3)\xi(\Psi_A-\Psi_V)+6u X_A^{(1)}-24u Y_A^{(1)}\right)\right.\nonumber \\
&\left.-2m_B^3(u-1)(\Psi_A-\Psi_V)+2m_B\Big((u-1)(q^2-\Sigma_{\xi})(\Psi_A-\Psi_V)\right.\nonumber\\
&\left.+3u(u\xi+\omega)\xi\left(4W^{(1)}-X_A^{(1)}+4Y_A^{(1)}\right)\Big)-2u(\omega+u\xi)(q^2-\Sigma_{\xi})(\Psi_A-\Psi_V)\right],\nonumber \\
&I_{3}^{(2)}(\omega,\xi,q^{2},\Omega)\nonumber \\ =&\ \frac{1}{m_B^6}({m_B}-\xi  u-\omega ) \left[m \left({m_B}^2 (-6 W^{(1)}-2
   {\bar X}_A^{(1)}+{X}_A^{(1)}+12 {\bar Y}_A^{(1)}-6 {Y}_A^{(1)})\right.\right.\nonumber\\
   &\left.\left.-12 {m_B}(W^{(2)}-4 Z^{(2)})+(\Sigma_{\xi}-{q^2}) (2
   {\bar X}_A^{(1)}-{X}_A^{(1)})\right)+{m_B}^3 \left(4 u {X}_A^{(1)}-2
   {\bar X}_A^{(1)}-{X}_A^{(1)}\right)\right.\nonumber\\
   &\left.+2 {m_B}^2 \left((\xi  u+\omega ) (-2 u
   {X}_A^{(1)}+2 W^{(1)}+{\bar X}_A^{(1)}+{X}_A^{(1)}+4 {\bar Y}_A^{(1)}+2
   {Y}_A^{(1)})-6 W^{(2)}+36 Z^{(2)}\right)\right.\nonumber\\
   &\left.+{m_B} \left({q^2} (-4 u{X}_A^{(1)}+2 {\bar X}_A^{(1)}+{X}_A^{(1)})+\Sigma_{\xi} \left(6 (4 u-1) W^{(1)}-4 u
   {X}_A^{(1)}+24 u {Y}_A^{(1)}+2 {\bar X}_A^{(1)}\right.\right.\right.\nonumber\\
   &\left.\left.\left.+{X}_A^{(1)}-6 (2
   {\bar Y}_A^{(1)}+{Y}_A^{(1)})\right)+12 W^{(2)} (\xi  u+\omega )\right)+2 \left(\Sigma_{\xi} \left(6
   W^{(2)}-(\xi  u+\omega ) ((12 u-1) W^{(1)}\right.\right.\right.\nonumber\\
   &\left.\left.\left.-2 (u{X}_A^{(1)}+{\bar Y}_A^{(1)})+(12 u-1) {Y}_A^{(1)}+{\bar X}_A^{(1)})\right)-{q^2}
   \left((\xi  u+\omega ) (-2 u {X}_A^{(1)}+W^{(1)}-{\bar X}_A^{(1)}\right.\right.\right.\nonumber\\
   &\left.\left.\left.+2
   {\bar Y}_A^{(1)}+{Y}_A^{(1)})+6 W^{(2)}\right)\right)\right],\nonumber \\
&J_{3}^{(2)}(\omega,\xi,q^{2},\Omega)\nonumber \\ =&\ \frac{1}{m_B^6}({m_B}-\xi  u-\omega ) \left[ {m_B}^2\left( m(2
   {\bar X}_A^{(1)}-{X}_A^{(1)})+2(\omega+u\xi)\left(W^{(1)}-{\bar X}_A^{(1)}\right.\right.\right.\nonumber\\
   &\left.\left.\left.+2{\bar Y}_A^{(1)}\right)+12 W^{(2)}\right)+2m_B\left(6\left(4mZ^{(2)}-mW^{(2)}+u\xi W^{(2)}+\omega W^{(2)}\right)\right.\right.\nonumber\\&\left.\left.+\Sigma_{\xi}\left((12u-3) \left(W^{(1)}+Y_A^{(1)}\right)+(1-4u) X_A^{(1)}-6 {\bar Y}_A^{(1)}+2{\bar X}_A^{(1)}\right)\right)\right.\nonumber \\
   &\left.-q^2\left(m\left(2{\bar X}_A^{(1)}-X_A^{(1)}\right)+2(\omega+u\xi)\left(W^{(1)}-{\bar X}_A^{(1)}+2{\bar Y}_A^{(1)}+Y_A^{(1)}\right)+12 W^{(2)}\right)\right.\nonumber \\&\left.+\Sigma_{\xi}\left(m\left(2{\bar X}_A^{(1)}-X_A^{(1)}\right)-2(\omega+u\xi)\left((12u-1)\left(W^{(1)}+Y_A^{(1)}\right)-4u X_A^{(1)}+{\bar X}_A^{(1)}\right.\right.\right.\nonumber\\&\left.\left.\left.-2{\bar Y}_A^{(1)}\right)+12 W^{(2)}\right)\right]~,\nonumber \\
& I_{3}^{(3)}(\omega,\xi,q^{2},\Omega)\nonumber \\ =&\  \frac{12}{m_B^{10}}(\omega+u\xi-m_B)^2\left[m_B(q^2-\Sigma_{\xi})\Sigma_{\xi} W^{(2)}+m_B^3 \Sigma_{\xi} \left(W^{(2)}-6 Z^{(2)}\right)\right.\nonumber\\&\left.+m m_B^2\left((\Sigma_{\xi}+q^2-m_B^2)W^{(2)}-6\Sigma_{\xi}Z^{(2)}\right)+(q^2-\Sigma_{\xi})^2(\omega+u\xi)W^{(2)}\right.\nonumber\\&\left.+4m_B^4(\omega+u\xi)Z^{(2)}-m_B^2(\omega+u\xi)\left((q^2+\Sigma_{\xi})W^{(2)}+2(2q^2-\Sigma_{\xi})Z^{(2)}\right)\right]~,\nonumber \\
& J_{3}^{(3)}(\omega,\xi,q^{2},\Omega)\nonumber \\ =&\  \frac{12}{m_B^{10}}(\omega+u\xi-m_B)^2\left[(m_B \Sigma_{\xi} (q^2-\Sigma_{\xi}-m_B^2) W^{(2)}+2 m m_B^2 \left(W^{(2)}-3 Z^{(2)}\right)\right.\nonumber\\&\left.+m_B^4 (\omega+u\xi)W^{(2)}+ (\omega+u\xi)(q^2-\Sigma_{\xi})^2 W^{(2)}-2m_B^2(\omega+u\xi)\left(q^2 W^{(2)}+\Sigma_{\xi} Z^{(2)}\right)\right]~,
\end{align}

where $\Sigma_0=\Sigma(\omega,0,q^{2},\Omega)$ and $\Sigma_{\xi}=\Sigma(\omega,\xi,q^{2},\Omega)$.

\end{appendix}

\end{document}